\documentclass[10pt,twocolumn,letterpaper]{article}

\usepackage{times}
\usepackage{epsfig}
\usepackage{graphicx}
\usepackage{amsmath}
\usepackage{amssymb}
\usepackage{subcaption}

\usepackage[noend]{algpseudocode}
\usepackage{algorithm}
\usepackage{url}
\usepackage{relsize}
\usepackage{xcolor,colortbl}
\usepackage{multirow}
\usepackage{wrapfig}
\usepackage{bbm}
\usepackage[labelfont=bf]{caption}
\usepackage{tabularx}

\usepackage[colorlinks = true,
            linkcolor = blue,
            urlcolor  = blue,
            citecolor = blue,
            anchorcolor = blue]{hyperref}

\begin{document}

\title{\vspace{-1cm}SneakPeek: \\
Interest Mining of Images based on User Interaction\vspace{-.25cm}}


\author{Daniyal Shahrokhian\thanks{Both authors contributed equally.} \hspace{20pt} Alejandro Vera de Juan\footnotemark[1]\\
\small{KTH Royal Institute of Technology}}

\maketitle

\begin{abstract}
\vspace{-.25cm}

\end{abstract}
Nowadays, eye tracking is the most used technology to detect areas of interest. This kind of technology requires specialized equipment recording user's eyes. In this paper, we propose \textit{SneakPeek}, a different approach to detect areas of interest on images displayed in web pages based on the zooming and panning actions of the users through the image.

We have validated our proposed solution with a group of test subjects that have performed a test in our on-line prototype. Being this the first iteration of the algorithm, we have found both good and bad results, depending on the type of image. In specific, \textit{SneakPeek} works best with medium/big objects in medium/big sized images. The reason behind it is the limitation on detection when smart-phone screens keep getting bigger and bigger.

\textit{SneakPeek} can be adapted to any website by simply adapting the controller interface for the specific case.

\section{Introduction}
\label{sect:introduction}

Interest tracking is a powerful tool when it comes to user experience testing. The fields of advertising, entertainment, packaging and web design have all benefited significantly from studying the visual behavior of the consumer.

\begin{figure}[!ht]
\centering
\includegraphics[width=0.45\textwidth]{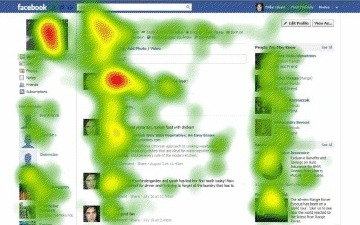}
\caption{One example of the applications of Interest Tracking: User Interaction (UI) design.}
\label{applications-interest-tracking}
\end{figure}

Eye tracking data is collected using either a remote or head-mounted ‘eye tracker’ connected to a computer. While there are many different types of non-intrusive eye trackers, they generally include two common components: a light source and a camera. The light source (usually infrared) is directed towards the eye. The camera tracks the reflection of the light source along with visible ocular features such as the pupil. This data is used to extrapolate the rotation of the eye and ultimately the direction of gaze. Additional information such as blink frequency and changes in pupil diameter are also detected by the eye tracker.

Eye Tracking can be an interesting approach, but lacks a general application in the common life-style of the users. This means that in current solutions, an eye-tracking device must be used on a test-subject, and a researcher must be, by their side, conducting the experiment. This limits the amount of data that can be retrieved.

\begin{figure}[!ht]
\centering
\includegraphics[width=0.35\textwidth]{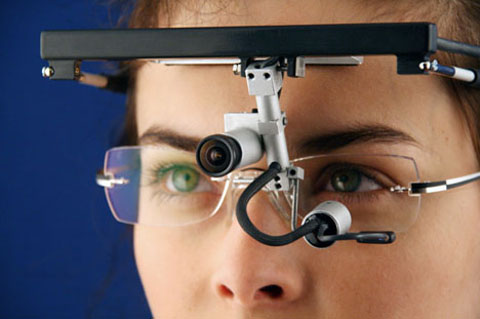}
\caption{Eye tracking device \cite{eye-tracking-article}.}
\label{eye-tracking-device}
\end{figure}

We want to simplify the way human interest measurements can be retrieved in images, so larger amounts of data can be gathered in a non-experimental environment. We also believe that the results of our approach overwhelm the results that can be achieved in an experimental setting, which biases the behavior of the test subjects.

Our goal is to validate if the way we are approaching interest area detection, using only user interface interaction, has an overall performance that makes it as suitable as the alternative that involves hands-on experimentation with, for example, eye-tracking devices.

As a case of use, our solution is very useful in web pages of clothing and accessory retailers, such as Zara, Aliexpress, H\&M, etc. These pages display images of their products and give the user the possibility to zoom and move through the image. These images usually display an outfit composed by multiple clothes. Implementing our proposed solution, these companies would understand in which of the clothes of the outfit users are more interested.

\subsection{Literature study}
\label{sect:framework}

This project builds on the idea that current approaches for detecting areas of interest consist of direct user experimentation, with technologies such as eye-tracking devices. Eye movements provide information about the location of areas of interest in an image  (Mackworth and Morandi \cite{Mackworth1967}; Just and Carpenter \cite{Just1976441}; Henderson and Hollingworth 1998 \cite{Henderson1998269}). Eye tracking devices can record the position and time an eye is looking at a point in the image,  more known as \textit{fixation time}.

In order to obtain the fixation time, some techniques have been developed. For example, Mackworth and Morandi \cite{Mackworth1967} propose to divide the image into a regular grid and count the time spent in each of the cells of the grid.

Some research has been performed based on this information. Santella and De Carlo \cite{Santella2004RobustCO} present an automatic data-driven method able to generate a representation  of viewer interest based on clustering  visual point-of-regard measurements into gazes (spatial clusters of successive fixations) and regions of interest.

Another example of clustering to extract regions of interest is the method used by Latimer \cite{Latimer1988}, which consists on creating a histogram of fixation durations on the image. Using this, it later clusters that histogram using k-means. 

Lastly, we have another clustering method in which eye-tracking is not taken into account. Instead of using user data to gather interest areas, G. Kim and A. Torralba \cite{kim_torralba_2009} propose an unsupervised version which introduces a fast and scalable alternating optimization technique to detect regions of interest (ROIs) in cluttered Web images without labels.

In order to compare areas of interest, we need to measure the distance between those areas. Huttenlocker \cite{Huttenlocher:1993:CIU:628305.628513} uses the Hausdorff distance to obtain the degree of resemblance between two objects in an image. The Hausdorff distance measures the degree of difference between two shapes. Shapes can be seen as a set of points, and according to the Hausdorff distance, two sets of points are close if every point in one set is close to some point of the other set.

Jaccard introduces in \cite{jaccard} the Jaccard index, more known as \textit{Jaccard similarity}. This measure allows to obtain the degree of similarity of two sets counting the number of elements in common and dividing it by the total number of elements between them. This measure can also be used with shapes if they are converted to a set of points inside those shapes.

After a deep research on literature, only A. Carlier \textit{et al}. \cite{axel_carlier_2010} have done something similar in the interaction aspect, with a completely different scope. They used pan and zoom measurements to determine which parts of a video were interesting for certain users, in order to crop the video to those sections. The purpose is to enhance user experience in devices with reduced screen size, such as smart phones.

\subsection{Hypotheses}
\label{sect:questions}

Our purpose is to test the hypothesis of whether areas of interest in images can be detected through simple interaction recordings (zoom, panning and time spent looking at a certain area).

\section{Method(s)}
\label{sec:method}

The project is based on the empirical method, because the only way to verify the proposed solution is through validation, using performance checking of the results of the experiments. 

Given that this is a quite novel approach compared to the current technologies, we pursue a one-iteration research: 
\begin{enumerate}
\item Develop a first version of the algorithm.
\item Gather data via user testing, and confirm whether our platform detects what they found interesting.
\item Analyze the results, and determine the problems of the algorithm. 
\item Propose an improvement on those weak points.
\end{enumerate}

After doing desktop research, we could not find datasets freely-available with records of the movements a subject does in an image (zoom, panning). Consequently, we needed to take care of generating this information through tests.

Subjects are shown a few images, and they are free to interact with them in any manner. In order to force the test subjects to zoom and move through the image, the images have a high resolution and are displayed in a small size. After this process, the same images are shown once again and the test subjects are asked to mark the areas of the image they considered most interesting.

Once we retrieved the data from all our test subjects, we analyzed the results comparing the output of our algorithm (a heat-map) with the areas of interest that the tests subjects marked as interesting. It is important to note that we have different types of data. On the one hand, the heat map will have continuous values in the interval [0,1]. On the other hand, the areas marked by the test subjects are binary (these areas are interesting and the rest of the image is not). Consequently, we need a way to compare both types of values.

In order to do this comparison, we have defined an \textit{interest threshold} to convert the continuous values to the binary values we have to compare with. Then, the Jaccard similarity is calculated by contrasting the total number of pixels above that threshold inside the heat-map and the pixels inside the areas marked by the user.

\subsection{Data retrieval}
We developed a test platform to perform tests with users. This test platform is based on a web server that provides a web page where the tests are performed. The test is composed by two phases displaying a set of images.

\paragraph*{Phase 1 - Free interaction with the image.}
The web page first displays an explicative message, giving the users some hints on the test and how to perform it. The message says: "\textit{Checkout these images. You can zoom and pan as you wish. We may ask you something about them later. When you finish, click Next.}".

The message was redacted trying to minimize the influence on the test subject while providing enough information to perform the test, as we needed the tests subjects to behave in the same way they behave when they see an image in a web page.

\begin{figure}[!ht]
\centering
\includegraphics[width=0.35\textwidth]{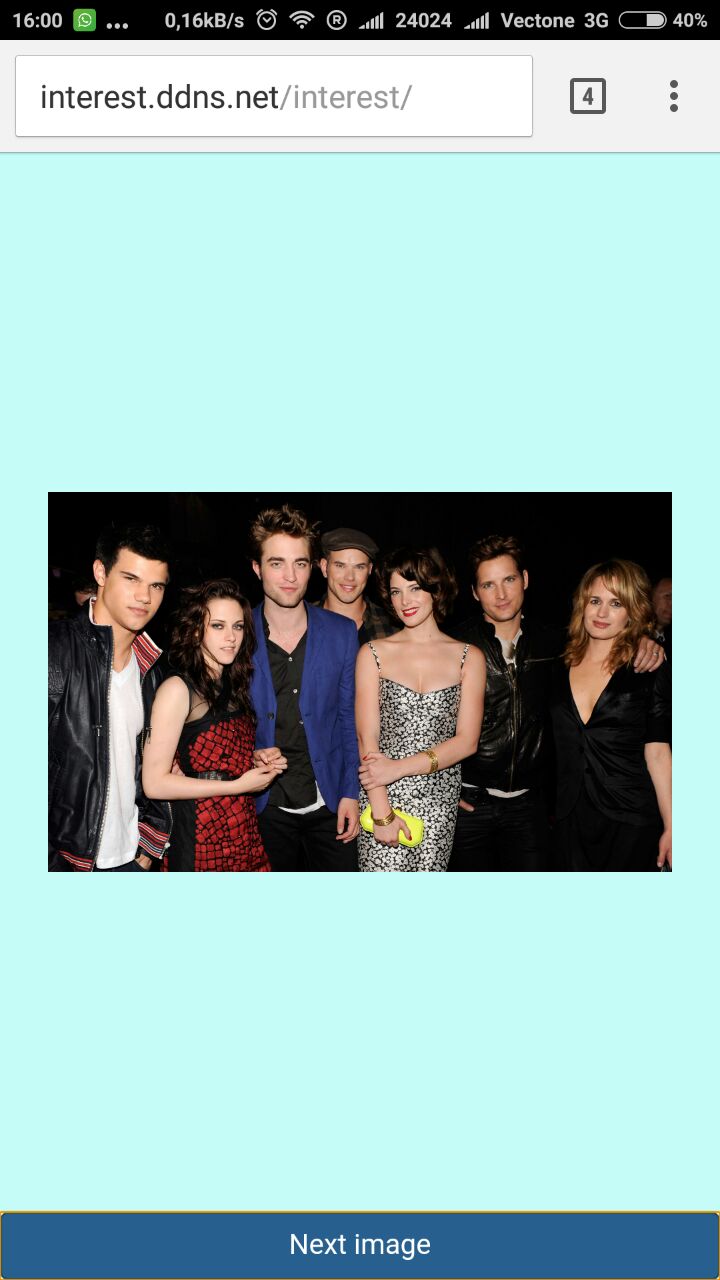}
\caption{Test Phase 1.}
\label{test-phase-1}
\end{figure}

After that, an image is displayed to the users so that they can zoom and pan it freely to check the most interesting parts of the image. Once they are finished, they can click in the next button, which will display the next image (see Figure \ref{test-phase-1}). 

In order to force the user to zoom and pan through the image, the high resolution images selected for this test are initially displayed in a small size.

\paragraph*{Phase 2 - Explicit selection of areas of interest.}
If all the test images have been displayed to the test subject, the test enters in a new phase with the intention to retrieve explicitly the areas of the image the test subject considers interesting. As in the previous phase, an explicative message is displayed to the user. This message says: "\textit{Now we need you to draw squares surrounding the interesting elements in the image. As in the previous phase, you can move and zoom to the areas you want and then click on the first button to enable the drawing mode (click again to disable). After that, select with your finger the interesting areas. If you want to undo your last action, use the second button. When you finish, click Next.}"

\begin{figure}[!ht]
\centering
\includegraphics[width=0.35\textwidth]{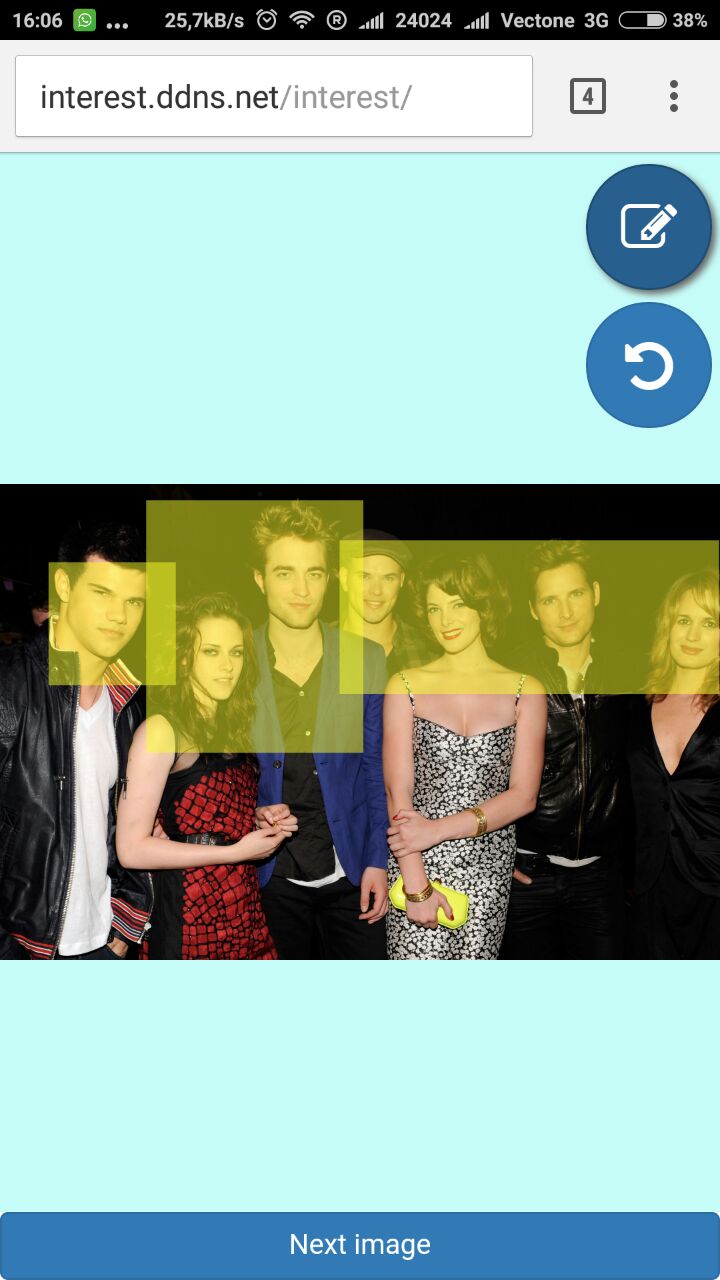}
\caption{Test Phase 2.}
\label{test-phase-2}
\end{figure}

After that, the same images used in phase 1 are displayed one by one to the test subject. They can move trough the image as in phase 1, but now with the ability to mark with squares the areas they consider are interesting in the image (see Figure \ref{test-phase-2}).

\section{The Platform} \label{sec-interest-platform}
\begin{figure}[!ht]
\centering
\includegraphics[width=0.3\textwidth]{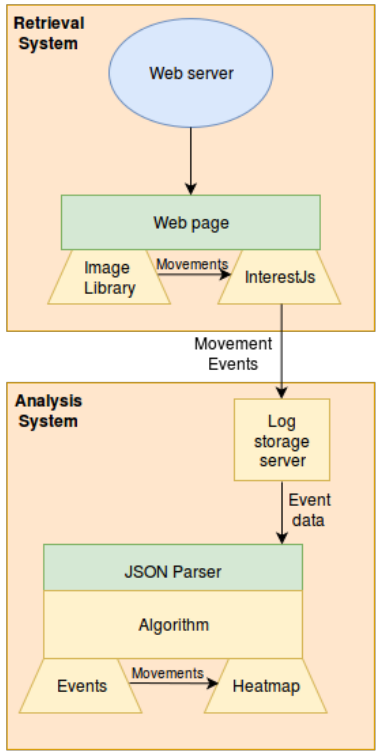}
\caption{Architecture of \textit{SneakPeek}.}
\label{interest-platform}
\end{figure}

In this section, we describe the proposed architecture (Figure \ref{interest-platform}) for a production platform to extract the interesting areas of images. 

We are using a client-server architecture. The client (Retrieval System) gathers and sends information regarding user actions, while the server (Analysis System) processes this information and generates the interest metrics. 

Auxiliary, we have also a Validation system, which is used during the analysis on the performance of our platform with human subjects. The communication between these components will be described when tackling each component.

\subsection{Retrieval system}
The Retrieval system is in charge of gathering interface information from the user. Zoom-able images are usually implemented in web pages with libraries, and the browser does not provide any support to this. These libraries, sometimes written ad-hoc, implement this zooming capabilities in heterogeneous ways. This makes it hard to find an automatic way to gather the interface events needed for our algorithm (zoom and pan).

Consequently, we decided to implement a Javascript client library (InterestJs) \cite{InterestJs} that allows web developers to easily integrate our system into their library to send the information through a REST API. This API provides methods to log the zoom and pan interface events, with information such as user ID, image ID, timestamp of the action, and bounding box of the new position in the image. Therefore, developers will only have to modify their controller methods for those events in their Image Library to add the call to the API.

\begin{figure}[!ht]
\centering
\includegraphics[width=0.3\textwidth]{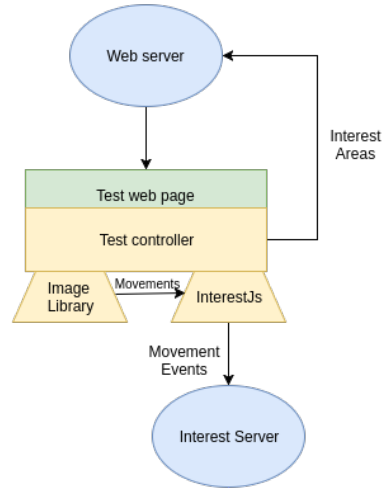}
\caption{Test platform in which the experiments are run.}
\label{test-platform-architecture}
\end{figure}

The system implemented \cite{webdemo, interestclient} to gather data from our test subjects extends this component. As it can be seen in Figure \ref{test-platform-architecture}, a Test controller has been added. This layer is responsible of executing both phases of the test and sending the data retrieved in phase 2 to the Web Server.

\subsection{Analysis system}
The Analysis system \cite{sneakpeek-server} is in charge of receiving the user-interaction data from the client, process this data with an algorithm, and generate a heat-map representing the interest of the user in different areas.

The Analysis System executes in the following way:

\begin{enumerate}
    \item The event handler receives a new interface event from a client.
    \item The event handler sends this data to the storage system.
    \item The storage system permanently saves this information.
    \item When a process requests the interest metric of an image, the algorithm is executed for each user that has interacted with the image.
    \item The results of all the users are normalized according to the maximum and minimum interest of each section of the image, and returned to the calling process.
\end{enumerate}

\begin{figure}[!ht]
\centering
\includegraphics[width=0.49\textwidth]{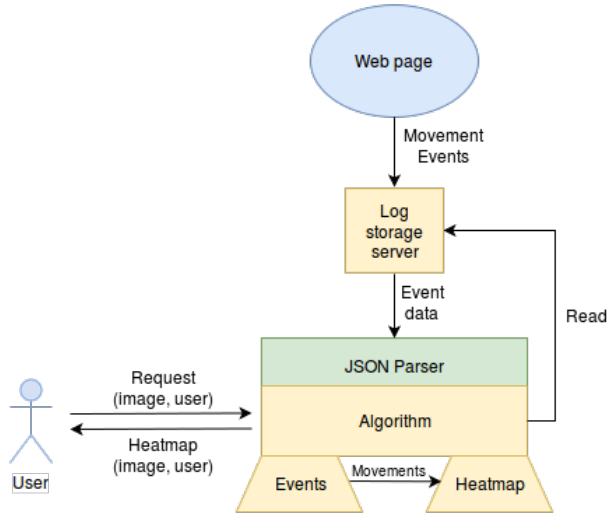}
\caption{Analysis System.}
\label{analysis-system}
\end{figure}

We used a simple JSON REST server that exposes all JSON files contained in a directory tree. The server listens to RESTful requests on a given port, and stores the information in the Data Storage.

The format of the events enables for stacking. Each event can be handled and incorporated into the system upon arrival. This behavior is typical of \textit{Log Structured Storage} --- that is, an append-only sequence of data entries.

Given the identifiers of the test, user and image involved in the transmission, we can isolate each action for each image and user easily. The only difference is that in \textit{SneakPeek} there is more than a single log. We have used a simple file hierarchy for this purpose: \textit{/test/image/user}.

Our API involves appending new data at the end of each file. Each file is indexed by [image, user], which implements native mutual exclusion (each file is accessed by a single user).

The complete algorithm for generating the heat-maps can be found in Algorithm \ref{alg:the_alg}. The overview on the work-flow of the algorithm is the following.

\begin{enumerate}
    \item Gather the event-log from certain (user,image).
    \item Generate a zero-interest heat-map with a size equal to the size of the image.
    \item Transform each event into an effect on the temporary heat-map.
    \begin{itemize}
        \item Screen-focus event: This is done by doing a counter on the time spent visualizing each pixel of the image. This is later normalized according to the minimum and maximum values. The main assumption is that the time spent at certain area is correlated to the user interest in such area.
        \item Zoom event: If the user dives-in/out of a section of the image, this event is recorded. We use a multiplier for the interest of the zoomed-in area based on the interest of the outer area, and the other way around. It is inverse to the area: the bigger the area, the less effect has the overall interest weight. We normalize this feature according to the size of the image (being the total area of the image equal to zero weight, and as we decrease this size, the weight increases linearly).
        \item Panning event: Fast movements across the image might mean lack of interest by the user. On the other hand, small movements around certain area of the image might mean interest of the scanned area.
    \end{itemize}
    \item Generate results: We have two different outputs. First, the heat-map of the interest areas across the image. Second, a deterministic number of areas that the algorithm found most relevant. In the current version, we use a threshold in the interest of each pixel. If the interest of such pixel is above average, we take it as relevant.
\end{enumerate}

\begin{algorithm}[tp]
\caption{SneakPeek algorithm}\label{sneakPeek}
\label{alg:the_alg}
\begin{algorithmic}[1]
\Procedure{Init}{image}
  \State $\textit{maxInterestVal} := 0$
  \State $n := \textit{image.width}$ 
  \State $m := \textit{image.height}$
  \State $\textit{heatMap} := \begin{bmatrix}
       0_{11} & 0_{12} & \cdots & 0_{1n}           \\[0.3em]
       0_{21} & 0_{22} & \cdots & 0_{2n}           \\[0.3em]
       \vdots & \vdots & \cdots & \vdots \\[0.3em]
       0_{m1} & 0_{m2} & \cdots & 0_{mn}
     \end{bmatrix}$
  \State $\textit{prevEvent} := \textit{None}$
\EndProcedure

\Procedure{\textit{AddEvent}}{\textit{newEvent}}
  \State $apply(getInterest(newEvent), prevEvent)$
  \State $\textit{prevEvent} := \textit{newEvent}$
\EndProcedure

\Procedure{Apply}{interest, event}
  \For{$\textit{all pixels (x,y) within the event area}$}
      \State $heatMap[x][y] := heatMap[x][y]$
      \State $ + interest$
      \State $\textit{maxInterestVal} :=$
      \State $max(\textit{maxInterestVal}, heatMap[x][y])$
    \EndFor
\EndProcedure

\Procedure{GetInterest}{event}
  \State $areaWidth := |prevEvent.upperLeftX $
  \State $- prevEvent.bottomRightX|$
  \State $areaHeight = |prevEvent.upperLeftY$ 
  \State $- prevEvent.bottomRightY|$    
  \State $timeDiff := event.time - prevEvent.time$
  \State $areaDiff := |image.width - areaWidth|$
  \State $+ |image.height - areaHeight|$
  \State \textbf{return} $\frac{areaDiff}{2} \cdot \textit{timeDiff}$
\EndProcedure

\Procedure{GetHeatMap}{}
\State $output := \begin{bmatrix}
       0_{11} & 0_{12} & \cdots & 0_{1n}           \\[0.3em]
       0_{21} & 0_{22} & \cdots & 0_{2n}           \\[0.3em]
       \vdots & \vdots & \cdots & \vdots \\[0.3em]
       0_{m1} & 0_{m2} & \cdots & 0_{mn}
     \end{bmatrix}$
     \For{$\textit{all pixels (x,y) in the output}$}
        \If{$heatMap[x][y] > 0$}
          \State $output[x][y] :=$
          \State $[normalize(heatMap[x][y])\cdot \frac{1}{2}))$
          \State $\textit{ and } (50\%\textit{ opacity}) ]$
        \Else 
          \State $output[x][y] := 0\%\textit{ opacity}$
        \EndIf
    \EndFor
    \State \textbf{return} output;
\EndProcedure

\end{algorithmic}
\end{algorithm}

\subsection{Validation System}

We have implemented a Test controller that is responsible of executing both phases of the test and sending the data retrieved in phase 2 to the Test Web Server. This server then stores this information in JSON format in a MySQL database.

\begin{figure}[!ht]
\centering
\includegraphics[width=0.3\textwidth]{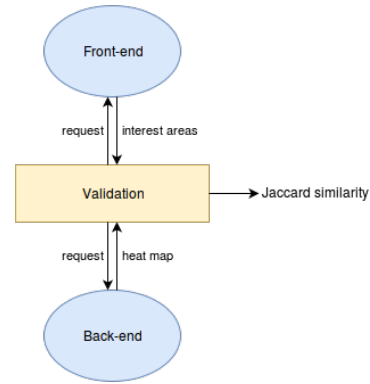}
\caption{Validation architecture.}
\label{val-arq}
\end{figure}

Once we have the user input, we proceed to validate the algorithm. First off, the algorithm is run with the event data gathered from the user. Then, the heatmap of these events is generated and validated against the marked areas by the user.

To measure the similarity between the output of the algorithm and the areas marked by the user, we define as output all pixels that have a higher interest value than the average over the whole image. Finally, we calculate the \textit{Jaccard similarity} between the areas marked by the user and the areas output of the algorithm.

\begin{figure}[!ht]
\centering
\includegraphics[width=0.45\textwidth]{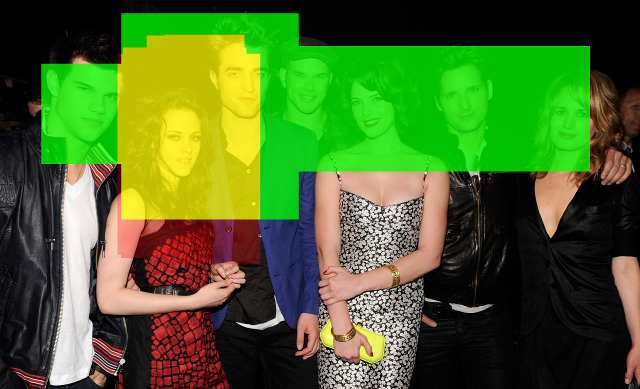}
\caption{Validation output of a given test.}
\label{test-phase-3}
\end{figure}

In Figure \ref{test-phase-3}, we can find an example of such validation. The red areas represent the heatmap output from \textit{SneakPeek}; The green ones represent what the user has marked as interesting; and the yellow ones represent the Jaccard similarity: areas in which the previous two overlap.

\section{Results and Analysis}
\label{sec:results}

We have gathered data of 34 different users for the first instance of \textit{SneakPeek}. We added a visualization tool to showcase the results of the tests, which includes the heat-map, the marked areas by the user and the intersection of both. 

In Figure \ref{barchart}, the results show variance depending on the image under test. In appendix \ref{sec:test_imgs}, we can find the images used in the experiment. The first four represent different number of objects, object sizes and image sizes. The last image was an easter egg in which the user was supposed to find Waldo.

\begin{figure*}[!ht]
\centering
\includegraphics[width=0.9\textwidth]{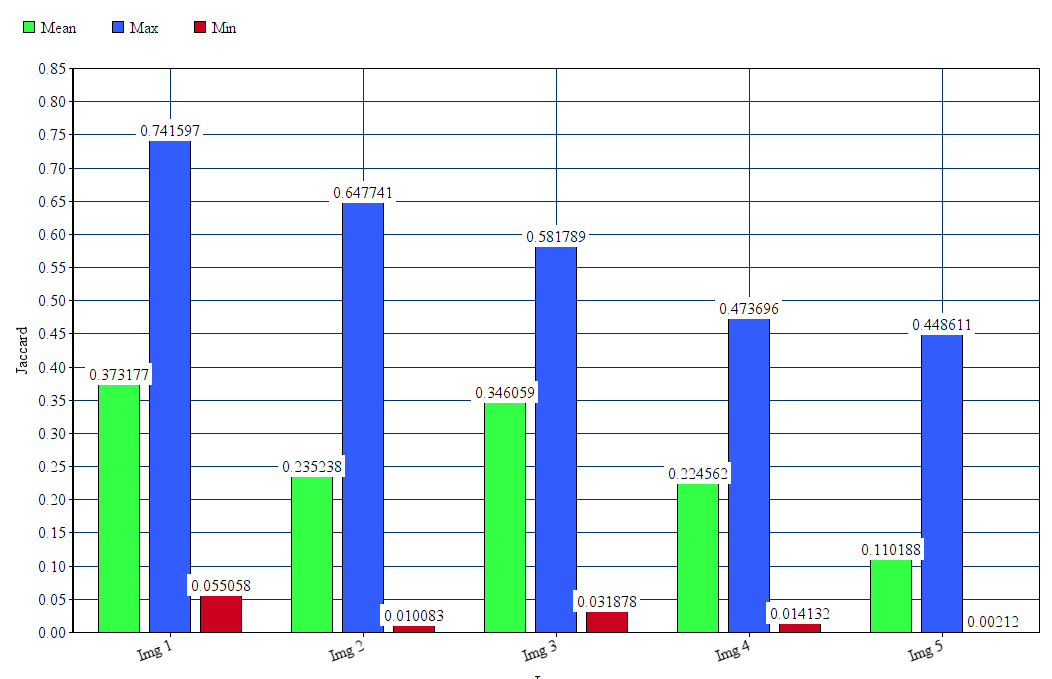}
\caption{Bar chart of the test results over different images.}
\label{barchart}
\end{figure*}

For instance, Image 1 recorded the best performance of the platform. This image reflects big objects in medium-sized images, which seem to work the best in \textit{SneakPeek}. There is a big problem in the experiments ran, and this is that in all of them, the minimum and maximum \textit{Jaccard similarity} differ widely from the average.

\begin{figure}[!ht]
\centering
\includegraphics[width=0.45\textwidth]{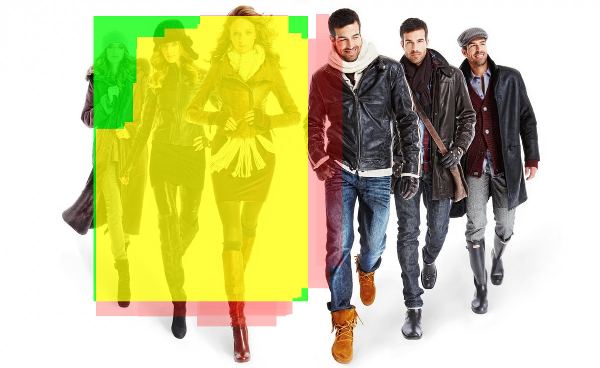}
\caption{Sample of the results achieved with Image 1.}
\label{good_test}
\end{figure}

In the case of Image 3, the results were not as good, since on average the algorithm seems to showcase a wider area of interest than the one the user points out. This can be tackled by increasing the threshold of which the algorithm classifies the area as interesting or not.

\begin{figure}[!ht]
\centering
\includegraphics[width=0.45\textwidth]{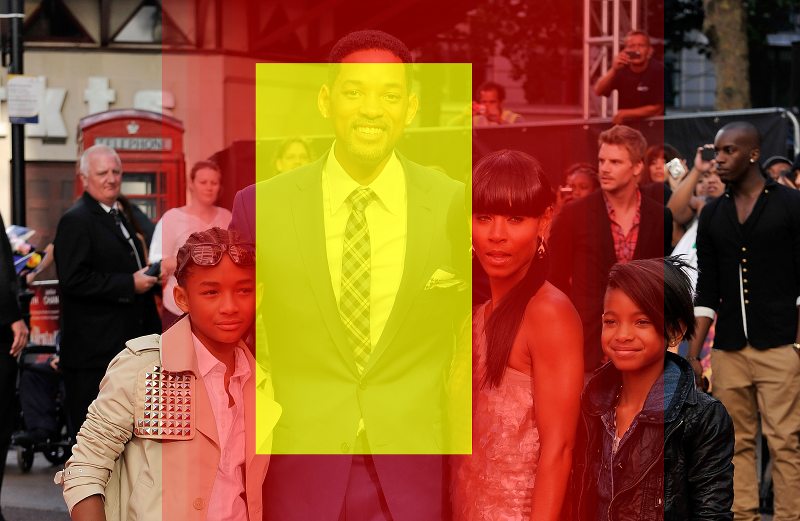}
\caption{Sample of the results achieved with Image 3.}
\label{normal_test}
\end{figure}

Last, in Image 4, some users achieved great results when they found the actors as interesting, while some others achieved poor results when being interested in a certain face or object.

\begin{figure}[!ht]
\centering
\includegraphics[width=0.45\textwidth]{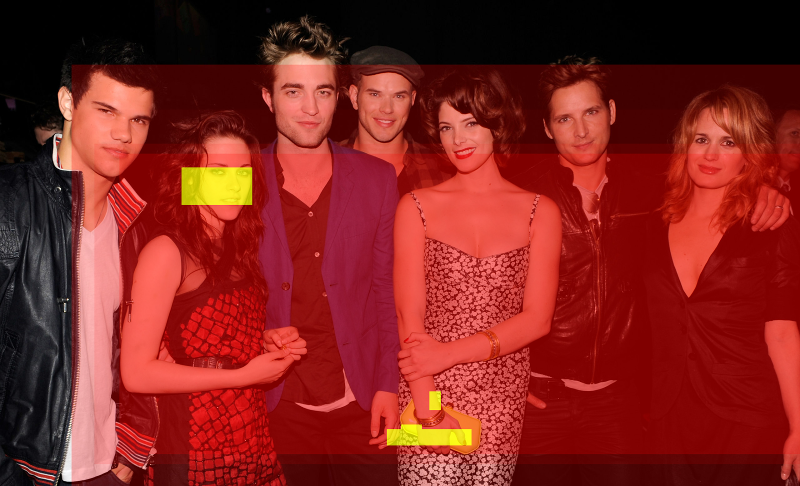}
\caption{Sample of the bad results in Image 4.}
\label{bad_test}
\end{figure}

In conclusion, the learning of the areas of interest seems to work really well with medium to big sized objects in the image, while tends to fall behind when small objects of interest are present.

\section{Discussion}
\label{sec:discussion}

\subsection{Future work}

This paper reflects a first iteration over the usage of interface metrics to record interest patterns from users. If we extended the scope of this paper further, a series of improvements could have been implemented upon release.

First of all, the amount of data gathered in the experiments represents only a small chunk of all the data that could be gathered if \textit{SneakPeek} was deployed in a commercial environment. Consumer-based web pages have many thousands of users continuously navigating through their products.

Second, some aspects of the front-end appear faulty. For instance, the fact that \textit{SneakPeek} records user interaction given a constrained time-bound leads to a non-optimized recording of the user interaction. For instance, if the user is looking at the position (0,0) at \textit{time=0 ms}, and moves to (0,6) at \textit{time=100ms}, it might be obvious that the platform should infer that the user passed through (0,3) at \textit{time=50ms}.

Third, if we think about it, the threshold that differentiates whether an area is interesting or not by the algorithm is an \textit{optimizable parameter}. After many trials, it is hard to come with a threshold that works the best for all test images. Through model training, a process widely known in the field of Machine Learning, we can use Linear Regression to minimize the cost (in this case, the Jaccard similarity) based on optimizing the threshold that determines whether an area is considered interesting.

\subsection{Sustainability and Ethics}
There are certain ethical challenges to our research. First and most important, there is the privacy and anonymity of the users of the platform. In our case, we have maintained the anonymity of our test subjects by storing all the data retrieved anonymously. Consequently, the data will not be related to that person. However, if an industry modifies this aspect of our system, it could potentially be used to create a profile of that person and know exactly what they like.

Furthermore, we believe our system is less intrusive than the current eye tracking technologies. To retrieve the same data that our system is designed to gather (from thousands of Internet users around the world), the users should accept to start their web camera so that the web page can record a video of their faces while they are navigating. It is obvious that most users would not accept this. However, in our case, we just have to keep a log of the actions users do through their screen interface.

Regarding sustainability, our system can allow producers understand what clients want. This is related to Goal 12 of the United Nations’ (UN) sustainable development goals for 2030 (UN17) \cite{olinger2015transforming}: \textit{"Ensure sustainable consumption and production patterns"}. With our system, less resources can be spent producing goods that the clients will not like. Consequently, the use of Earth's resources can be optimized, and less materials and energy will be wasted. 

In addition, this system will enhance relations between producers and consumers by helping customers get what they want, so that they will not spend that much time looking for it.

\bibliography{II2202-report}
\bibliographystyle{myIEEEtran}

\end{document}